\documentclass[reprint,amsmath,amssymb,aps,prl,floatfix,superscriptaddress,twocolumn]{revtex4-2}
\usepackage[utf8]{inputenc}
\usepackage[T1]{fontenc}
\usepackage[english]{babel}
\usepackage{enumerate}
\usepackage{amsfonts}
\usepackage{booktabs}
\usepackage{stackengine}
\usepackage[normal]{subfigure}
\usepackage{floatflt}
\usepackage{listings}
\usepackage{graphicx}
\usepackage{amsmath}
\usepackage{csquotes}
\usepackage{hyperref}\hypersetup{colorlinks=true, allcolors=blue}
\usepackage{enumitem}
\usepackage{multirow}
\usepackage{soul}
\usepackage{mathtools}
\usepackage{orcidlink}
\usepackage{amsthm}
\usepackage{amssymb}
\usepackage{physics}
\usepackage{wrapfig}
\usepackage{dsfont}
\usepackage{comment}
\usepackage[title]{appendix}
\usepackage{xcolor}
\usepackage[normalem]{ulem}
\usepackage{pict2e}
\usepackage{relsize}
\usepackage{tikz}
\usetikzlibrary{positioning}
\usepackage{balance}

\usepackage{tgtermes}
\usepackage{newtxmath}

\DeclareMathAlphabet{\mathbb}{U}{msb}{m}{n}
\DeclareMathAlphabet{\mathfrak}{U}{euf}{m}{n}

\providecommand*{\unit}[1]{\,\ifmmode
\mathrm{\,#1}\else\textup{#1}\fi}
\newcommand{\Md}[1]{M_{#1}(\mathbb{C})}
      
\newcommand{\normt}[1]{\norm{#1}_1}
\newcommand{\bs}[1]{\boldsymbol{#1}}
\newcommand{\mc}[1]{\mathcal{#1}}

\newcommand{\orcidauthorFB}{0000-0002-0712-2057} 
\newcommand{\orcidauthorGN}{0009-0006-3232-1222} 

\theoremstyle{plain}

\theoremstyle{plain}

\begin{document}
	\title{Quantum Dynamical Entropy and Dissipative Information Flows}
	\author{Giovanni Nichele\,\orcidlink{\orcidauthorGN}}
	\email{gnichele@umk.pl}
	\affiliation{Institute of Physics, Faculty of Physics, Astronomy and Informatics, Nicolaus Copernicus University,  Grudziądzka 5/7, 87-100 Toruń, Poland}
	\affiliation{Dipartimento di Fisica, Università degli Studi di Trieste, I-34151 Trieste, Italy}
	\affiliation{Istituto Nazionale di Fisica Nucleare, Sezione di Trieste, I-34151 Trieste, Italy}
	
	\author{Fabio Benatti\,\orcidlink{\orcidauthorFB}}
	\affiliation{Dipartimento di Fisica, Università degli Studi di Trieste, I-34151 Trieste, Italy}
	\affiliation{Istituto Nazionale di Fisica Nucleare, Sezione di Trieste, I-34151 Trieste, Italy}
	\date{\today} 

\begin{abstract}
The Alicki–Lindblad–Fannes dynamical (ALF) entropy measures the rate at which new information is gathered about a quantum system by inspecting its long-time evolution. After extending the ALF entropy  to open quantum dynamics, we propose it as a measure of back-flow of information from the environment. We show that the behavior of the \emph{open-system ALF entropy} is a stronger witness than  existing approaches based only on the open system reduced dynamics. In the case of a qubit collisionally coupled to a classical spin chain, we obtain an exact expression for the open-system ALF entropy explicitly depending on the environment correlations. An extreme case shows how the information flow from environment to system corresponds to vanishing entropy production as for reversible finite quantum systems.
\end{abstract}

\maketitle

In recent years, much efforts have been devoted to the study of open quantum systems beyond the so-called Markovian regime, namely, when the dissipative dynamics of the system of interest is affected by non-negligible memory effects~\cite{ChrusReview22,LiLi,RivasHuelgaPlenio}.
Often, these effects 
are already witnessed at the level of the reduced dynamics of the system: we will consider it to be described, in the Schrödinger picture, by a one-parameter, discrete-time family of completely positive and trace-preserving (CPTP) maps $\Lambda^\ddag_{n}, n \ge0$. Striking features of non-Markovianity 
include, for example, revivals in time of distances or, more in general, of quantum divergences~\cite{BLP,Megier2021entropic,HolevoSkew, VacchiniBook2024,Settimo2022}. Such effects are usually interpreted as back-flow of information that has been temporarily stored outside the system and becomes re-injected into it at later times. On the other hand, the reduced dynamics by itself is neither sufficient to witness the totality of non-Markovian effects~\cite{MilzCPddoesntmean}, nor to capture
the physical roots  of the information flows underlying the memory mechanisms~\cite{FBGN_PhysicaScr}. 
 Classically, Markovianity concerns the multi-time correlation functions of a stochastic process,
of which the reduced dynamics is only a one-time marginal. 
The attempts at an exhaustive quantum extension of the notion of Markovianity stem back to the pioneering works~\cite{AccardiFrigerioLewis,Accarditopics,Lindblad1979}. Among the most recent proposals in this direction, particularly prominent ones include the so-called process tensor formalism~\cite{MilzModi_tutorial,ProcessTensorPRL,PollockModiPater2018}, quantum combs (see~\cite{Chiribella08,milz2020non} and references therein), conditional past-future independence~\cite{Budini2018quantum,BudiniOPvdNONOP} and temporal entanglement (see the recent work~\cite{vilkoviskiy2025temporal}). The common idea in most of them is to map multi-time correlations into spatial ones encoded in a quantum many-body state.
The same idea also underpinned the construction of 
quantum dynamical entropy,
namely a quantum generalization of Kolmogorov-Sinai (KS) entropy~\cite{Petersen_1983}, as developed by Alicki and Fannes~\cite{AF1994defining}, building on earlier work by Lindblad~\cite{Lindblad88}. Although Lindblad’s original interest concerned irreversible processes~\cite{Lindblad1979}, most subsequent applications of the Alicki-Lindblad-Fannes (ALF) entropy have exclusively focused on the reversible evolution of many-body quantum systems and their possibly chaotic behavior~\cite{AlickiFannesBook}.
In this Letter,
1) we instead exploit the multi-time statistics of open systems to extend the ALF entropy to dissipative evolutions and 2) we use it to entropically characterize system-environment information exchange in the presence of memory effects. Such a proposal fundamentally differs from approaches based solely on the reduced dynamics, as it takes into account the full multi-time statistics of the process. For this reason, as we shall see, it can capture memory effects that cannot be assessed by inspecting the reduced dynamics alone.
 The multi-time correlation functions, central to the construction of the ALF entropy, are gathered by repeated measurements on the system of interest. Clearly, handling correlation functions requires sufficient control of the system-environment interactions.
In this Letter, we investigate the behavior of the ALF entropy in a concrete dissipative setting by  focusing on the so-called collisional models~\cite{CMreview}.
These latter prove to be particularly convenient to describe interactions between an open system and a genuinely infinite environment, allowing the transition from the Markovian regime to the presence of memory effects. 
While the proposed approach extends well beyond collisional models, here we provide a self-contained account of the main methods and results. A more general treatment, including full technical derivations and further discussion, is given in the companion paper~\cite{PRA}.
Consider a finite open $d$-level system $S$, whose  
environment is provided by an infinite spin chain. Local observables $A_E^{[-a,b]}$ of the chain are supported by finite intervals $[-a,b]$ and belong to strictly local matrix algebras 
$M^{[-a,b]}_D(\mathbb{C})=\bigotimes_{k=-a}^b M^{(k)}_D(\mathbb{C})$. 
The environment is endowed with a state $\omega_E$ 
specified by a consistent set of density matrices describing the expectations of local observables
$
\omega_E\left(A_E^{[-a,b]}\right)=\Tr(\rho_E^{[-a,b]}A_E^{[-a,b]})
$;
consistency means that  $\Tr_{b+1}\rho_E^{[-a,b+1]}=\rho_E^{[-a,b]}$.
Furthermore, the environment state is assumed to be shift-invariant:  $\omega_E=\omega_E\circ\sigma_E$, $\sigma_E$  being the right shift on the chain: 
$\sigma_E[A_E^{[a,b]}]=A_E^{[a+1,b+1]}$.
In terms of local density matrices, shift-invariance is equivalent to the condition $\Tr_{-a}\rho_E^{[-a,b]}=\rho_E^{[-a,b-1]}$. The system state $\omega_S$ is instead described by a full-rank density matrix $\rho_S>0$, $\omega_S(X_S)=\Tr(\rho_SX_S)$, $X_S\in\Md{d}$. 
The $S+E$ collisional dynamics is then constructed as follows. Consider  a unitary map $\Phi$, acting non-trivially only  on the observables of $S$ and of the chain $0$-th site,
$\Phi[X_S\otimes X_E^{(0)}]=U^\dag X_S\otimes X_E^{(0)} U$.
The $S+E$ one-step Heisenberg dynamics is then given by composing $\Phi$ with the shift $\sigma_E$ on the environment chain,  
$	\Theta:= ({\rm id}_S\otimes\sigma_E)\circ\Phi
$ (see Figure~\ref{fig:collisionalfigure}(a)). The reversible dynamics at discrete-time $n$ is finally given by the group $\Theta_n=\Theta^n$, $n\in\mathbb{N}$, which we assume leaves the reference state invariant: $\omega_S\otimes \omega_E\circ\Theta=\omega_S\otimes \omega_E$.
Since the microscopic dynamics of the environment is hardly controllable, the practically accessible correlation functions are those involving operators $X_a$ of the open system~$S$ only. Concretely, we shall consider positive-operator valued measurements (POVM) $\mc{X}=\{X_a\otimes \mathds{1}_E\}_{a=1}^{\abs{\mc{X}}}$ based on operators 
such that $\sum_{a=1}^{\abs{\mc{X}}} X_a^\dagger X_a=\mathds{1}_d$.
Following a typical strategy of classical ergodic theory~\cite{Petersen_1983}, 
one  construct a quantum-like symbolic dynamics for the open system obtained through repeated observations. To this end,  
 let
\begin{equation}\label{multi-notation}
	{X}_{\bs{a}}^{(n)}:=\Theta_{n-1}[X_{a_{n-1}}\otimes\mathds{1}_E] \cdots\Theta_1[X_{a_1}\otimes\mathds{1}_E]X_{a_0}\otimes\mathds{1}_E 
	\,,
\end{equation}
where $\bs{a}:=a_0 \dots a_{n-1}$, and define the  $\abs{\mc{X}}^n\times \abs{\mc{X}}^n$ matrix whose entries are the multi-time correlation functions:
\begin{equation}\label{system_cmatrix}
	\left(\rho_S\big[\mathcal{X}^{(n)}\big]\right)_{\bs{a},\bs{b}}
	=\omega_S\otimes \omega_E\left(X_{\bs{b}}^{(n)\dagger} X_{\bs{a}}^{(n)}\right) .
\end{equation}
The matrix $\rho_S\big[\mathcal{X}^{(n)}\big]$ is positive with unit trace: we shall thus call it \emph{coarse-grained density matrix}. In contrast with the classical partitioning of the phase space that does not alter the classical states nor perturb the dynamics, 
POVM measurements affect quantum states and interfere with their time-evolution.  We can associate to the density matrix~\eqref{system_cmatrix} an entropy rate
\begin{equation}
	\label{ALFrate}
	\mathfrak{h}_S(\Theta,\mc{X}):=
	\limsup_{n} \frac{1}{n} S\left(\rho_S\left[\mc{X}^{(n)}\right]\right), 
\end{equation}
where $S(\rho)=-\Tr\rho \log \rho$ is the von Neumann entropy. One then maximizes~\eqref{ALFrate} over the all admissible POVMs so to obtain the \emph{open-system ALF entropy}:
\begin{equation}
	\mathfrak{h}_S(\Theta)\quad \ :=
	\sup_{\mc{X}\,\subseteq\,\Md{d} \otimes \mathds{1}_E}\mathfrak{h}_{S}(\Theta, \mathcal{X})\, .
	\label{ALFopen}
\end{equation}
A physical interpretation of the above quantity follows from a joint state purification of the compound system $S+E$, known mathematically as GNS representation (see~\cite{AlickiFannesBook}  and  the companion paper~\cite{PRA}). There, 
joint observables are represented as bounded operators on a suitable Hilbert space $\mc{H}_S\otimes\mc{H}_E$.
The joint purified state can be taken of the factor form $\ket{\Omega_{SE}}:=\ket{\sqrt{\rho_S}\otimes \Omega_E}$, while observables of the system are represented by tensor products $X_S\otimes \mathds{1}_d$. Furthermore, the joint collisional evolution can be implemented by a unitary operator $U_{\Theta}^\dag$ that leaves $\ket{\Omega_{SE}}$ invariant.
Then, remarkably,~\eqref{system_cmatrix} shares the same von Neumann entropy of 
\begin{equation}\label{GNS_matrix}
	 \Big(\mathbb{U}_{\Theta}^\ddag\circ \big(\mathbb{X}^\ddag\otimes\mathrm{id}_d\otimes\mathrm{id}_E\big)\Big)^n\, \![\ketbra{\Omega_{SE}}] \,, 
\end{equation}
where $\mathbb{X}^\ddag[\bullet]=\sum_a X_a \bullet X_a^\dag$ is the  map describing the POVM and $\mathbb{U}_\Theta^\ddag[\bullet]={U}_\Theta\bullet{U}_\Theta^\dag$. 
Hence, by the introduction of an ancilla,~\eqref{GNS_matrix} allows one to interpret the open-system ALF entropy as the optimal  information rate gathered about the open dynamics by subsequent measurements intertwining the unitary evolution, similarly as for the classical KS entropy~\cite{Petersen_1983, Alicki2002}.
If the quantum system $S$ is finite and closed, namely uncoupled to the environment so that $\Theta=\Theta_S\otimes \Theta_E$, one proves~\cite{AlickiFannesBook} that the entropy of~\eqref{system_cmatrix} is bounded from above by
$	S\left(\rho_S\left[\mc{X}^{(n)}\right]\right) \le 2 \log(d)\,.
$
The r.h.s, in particular, does not scale with $n$; dividing both sides by $n$ and taking $\limsup$ yields:
\begin{equation}\label{closedALF}
	\mathfrak{h}_S(\Theta)=0 \quad\hbox{for a closed, finite system}.
\end{equation}
In general, vanishing entropy rates signal that, in the long run, no uncertainty is left about the outcomes of the next measurement or that 
no  new information is gathered through it.\par
In the presence of a genuine coupling with the environment, the density matrix~\eqref{system_cmatrix} is rarely under analytical control. Luckily, it has a closed form with respect to the special POVM
$
	\mc{F}:=\{F_{a,a'}	\otimes \mathds{1}_E\}_{a,a'=1}^d$, 
$
F_{a,a'}=\sqrt{r_a} \ketbra{r_a}{r_{a'}},
$
where $r_a>0, \ket{r_a}$ are the eigenvalues, respectively, eigenvectors of $\rho_S$. 
To this end, consider an auxiliary quantum spin chain $M_d^{\mathbb{Z}}(\mathbb{C})=\bigotimes_{k=-\infty}^{+\infty} \Md{d}$ consisting of infinitely many copies of our original $d$-level open system $S$ . We endow the auxiliary chain
  with its own shift operator $\sigma_S$. 
At each tick of time,  the  dynamics $\Theta$ acts non-trivially  only on the 0-th copy $M_d^{(0)}(\mathbb{C})$ of the system $S$ and on the environment $E$ (see Figure~\ref{fig:collisionalfigure}(b)).  Based on the above construction, we introduce the \emph{correlations maps}:
\begin{equation}\label{mapTn}
	\mathbb{T}_n\bigg[\bigotimes_{j=0}^{n-1} A_j\bigg]:=\omega_E\bigg( (\Theta\circ (\sigma_S\otimes \mathrm{id}_E))^n\Big[\bigotimes_{k=0}^{n-1} A_{k}^{(-n+k)}\otimes\mathds{1}_E\Big]\bigg).
\end{equation}
Here, $n$ actions of the shift $\sigma_S$ push localized operators $n$ sites to the right on the system chain $M_d^{\mathbb{Z}}(\mathbb{C})$.  The  shift is interwtined by the action of $\Theta$ only on observables of $M_d^{(0)}(\mathbb{C})$ and on those of the environment. Finally, the map $\mathbb{T}_n$ in the Heisenberg picture is obtained by  eliminating the degrees of freedom of the environment through the partial expectation $\omega_E$.
\begin{figure}[t!]
	\begin{tikzpicture}[scale=0.89, transform shape, baseline=(current bounding box.center),
		every node/.style={font=\normalsize},
		tensor/.style={inner sep=1.1pt},
		graybox/.style={fill=purple!10, rounded corners=3pt, inner sep=2.6pt},
		purplebox/.style={fill=purple!10, inner xsep=10pt,   
			inner ysep=20pt, rounded corners=2.2pt,minimum height=2.5em},
		]

		\node[label] (int) at (-12.2,1.2) {\small{(a)}} ;
		\node[label] (phi) at (-10.3,1.2) {$\boxed{\Theta=(\mathrm{id}_S\otimes\sigma_E)\circ\Phi}$};

		\node[tensor] (left1) at (-10.2,-0.3){$\cdots 
			\otimes
			M_D^{(-a)} \otimes M_D^{(-a+1)} \otimes\cdots \otimes$};

		\node[graybox] (ASmid) at (-7.45,0.2){
			$
			\begin{aligned}
				&\ M_d\\
				&	\ 	\, \ \otimes \\
				& M_D^{(0)}
			\end{aligned}
			$
		};

		\node[tensor, right=.9cm of left1] (right1) {$\otimes\, 
			\cdots \otimes M_D^{(b)} \!\otimes M_D^{(b+1)} \otimes
			\cdots
			$
		};

	\end{tikzpicture}
	\vskip+2mm
	\hskip-1mm
	\begin{tikzpicture}[scale=0.88, transform shape, baseline=(current bounding box.center),
		every node/.style={font=\normalsize},
		tensor/.style={inner sep=1.1pt},
		graybox/.style={fill=blue!10, rounded corners=3pt, inner sep=2.6pt},
		purplebox/.style={fill=blue!10, inner xsep=10pt,   
			inner ysep=20pt, rounded corners=2.2pt,minimum height=2.5em},
		]

		\node[label] (int) at (-12.2,1.2) {\small{(b)}} ;
		\node[label] (phi) at (-10.7,1.2) {$\boxed{\Theta \circ (\sigma_S\otimes \mathrm{id}_E)}$};

		\node[tensor] (left1) at (-10.3,0.2){$\cdots 
			\otimes
			M_d^{(-a)} \otimes M_d^{(-a+1)}  \otimes\cdots \otimes$};

		\node[graybox] (ASmid) at (-7.45,-0.3){
			$
			\begin{aligned}
				&M_d^{(0)}\\
				&\ 	\, \ \otimes \\
				& M_D^{(0)}
			\end{aligned}
			$
		};

		\node[tensor, right=1cm of left1] (right1) {$\otimes\, 
			\cdots \otimes M_d^{(b)} \!\otimes M_d^{(b+1)} \otimes
			\cdots
			$
		};

		\node[tensor] (right1)  at (-4.9,-.9) {$\otimes\, 
			\cdots \otimes M_D^{(b)} \!\otimes M_D^{(b+1)} \otimes
			\cdots
			$
		};
		
		\node[tensor] (left2)  at (-10.3,-.9) {$  
			\cdots \otimes  M_D^{(-a)} \otimes M_D^{(-a+1)} \otimes
			\cdots\otimes
			$
		};

	\end{tikzpicture}
	\caption{(a) One step $\Theta$ of the Heisenberg collisional dynamics: $\Phi$ jointly acts on the system observables in $M_d\equiv M_d(\mathbb{C})$ and on those of the $0$-th site of the chain. (b) One step of the dynamics for the open quantum chain $M_d^{\mathbb{Z}}(\mathbb{C})$: at each tick of time, the system copy at the 0-th site interacts with the environment $E$, here represented as an infinite spin chain.}
	\label{fig:collisionalfigure}
\end{figure}
Then, up to rearrangement of tensor factors, one has
\begin{equation}\label{thm1_1}
	\rho_S\left[\mc{F}^{(n+1)}\right] =\rho_S\otimes \rho_S \otimes  \left(\mathbb{T}_n^\ddag \otimes \mathrm{id} \left[\ketbra{\sqrt{\rho_S^{\otimes n}}}\right]\right),
\end{equation}
where $\mathbb{T}_n^\ddag$ is the CPTP Schrödinger map corresponding to $\mathbb{T}_n$
and $\ket{\sqrt{\rho}}=\sum_a\sqrt{r_a}\ket{r_a}\otimes\ket{r_a}$ denotes the standard purification of $\rho=\sum_ar_a\,\ketbra{r_a}{r_a}$.
The special  POVM $\mathcal{F}$ is not necessarily the one achieving the supremum in~\eqref{ALFopen}. In particular, it preserves $\rho_S$, namely
	$\sum_{a,a'=1}^d F_{a,a'}\,\rho_S\, F^\dag_{a,a'}=\rho_S$. The supremum in~\eqref{ALFopen} is instead taken on all possible system POVMs,  not just those preserving  $\rho_S$. Nevertheless, as we shall see,~\eqref{thm1_1} is  important per se since it allows to identify  Markovianity  in terms of entropy rates. Indeed, when the multi-time statistics of the process is available, the most appropriate notion of Markovianity is the so-called Quantum Regression (QR) regime~\cite{Lax,ChrusReview22,LonigroChrusc1}.  
Namely, an open dynamics is called QR-Markovian when the system multi-time correlation functions, that is the coarse-grained density matrix~\eqref{system_cmatrix}, can be reconstructed from the reduced dynamics propagators:
\begin{equation}
		\left(\rho_S\left[\mc{X}^{(n)}\right]\right)_{\bs{a},\bs{b}}\!\!\!=\omega_S\left( \mathbb{X}_{a_0,b_0} \circ \Lambda\circ  \cdots\circ\Lambda \circ \mathbb{X}_{a_{n-1},b_{n-1}} [\mathds{1}_S]  
		\right),\label{QR}
\end{equation}
with $\mathbb{X}_{a_0,b_0}[\,\bullet\,]:={X}_{b_0}^\dagger \bullet {X}_{a_0} $ and where the (Heisenberg) dynamical CP identity preserving maps at time-step $n$, $\Lambda_n$, form a discrete semi-group: $\Lambda_n=\Lambda^n$ (see~\cite{PRA} for the QR formula in more general settings).
As already noted by Lindblad in~\cite{Lindblad1979} and   
proved in the companion paper~\cite{PRA}, it turns out that 
QR-Markovianity is equivalent to factorizing  correlation maps, $\mathbb{T}_n=\bigotimes_{j=0}^{n-1} \Lambda$. 
From this fact, the QR regime can be uniquely identified by entropic ALF rates. To this end, we introduce the non-negative quantity
\begin{equation}\label{ineq_entropy}
	\Delta \mathfrak{h}_S(\Theta,\mathcal{F})
	:= S\!\left(\Lambda^\ddag \otimes \mathrm{id}_d\left[\ketbra{\sqrt{\rho_S}}\right]\right)
	-\mathfrak{h}_S(\Theta,\mathcal{F})\,,
\end{equation}
which vanishes if and only if QR-Markovianity holds (see~\cite{PRA} for the proof). The asymptotic entropy rate quantifies the information provided by repeated inspections of the open dynamics;  if  such rate diminishes with respect to its value in the QR-Markovian regime, it means that the dynamics has become less unpredictable.  We can then interpret~\eqref{ineq_entropy} as a measure of information flowing back from the environment to the system that, unlike other proposals, relies on the full multi-time statistics.  \par
In order to highlight the use of the  open-system ALF entropy beyond the Markovian regime, we compute it analytically for a qubit system collisionally interacting with a classical stationary chain and show that it can even vanish. The collisional environment consists of  a chain, each site carrying  a same commutative, that is diagonal, matrix algebra, spanned by $1$-dimensional orthogonal projections $\{\Pi_i\}_{i=0}^{3}$, $\sum_{k=0}^{3}\Pi_k=\mathds{1}_4$.
Furthermore, the shift-invariant state of the chain corresponds to diagonal local density matrices 
$	\rho_E^{[-a,b]}=\sum_{\bs{i}_{[-a,b]}}p_{\bs{i}_{[-a,b]}}\,\Pi_{\bs{i}_{[-a,b]}}^{[-a,b]}\, ,
$
where $\Pi^{[-a,b]}_{\bs{i}_{[-a,b]}}:=\bigotimes_{k=-a}^{b} \Pi_{i_k}^{(k)}$ and $\pi_{[-a,b]} :=\{ p_{\bs{i}_{[-a,b]}}\}$ is a stationary probability distribution.
The reversible interaction is taken as 
$	\Phi[X_S\otimes A_{i_0}^{(0)}]=\sum_{k=0}^{3}\phi_{k}[X_S]\otimes \Pi_k^{(0)} A_{i_0}^{(0)} \Pi_k^{(0)}$,
where $\phi_k[X_S]:=\sigma_k\, X_S\, \sigma_k	\nonumber$, $\sigma_0=\mathds{1}_S$ and $\sigma_k$, $k=1,2,3$ are the Pauli matrices. Such a controlled-unitary interaction, typical of many collisional models~\cite{RybarFilippovZiman,BernardesCM,BernardesExperimentalObservationWeak2015}, has been extensively discussed in the open-systems literature~\cite{backer_localdisclosure,BAV_2018,buscemi2025causal}. Despite the environment  being entirely classical, its coupling to the open quantum system can still generate memory effects with no classical analogue, such as the so-called super-activation of back-flow of information~\cite{FBGN_PhysicaScr}.
Then, let the reference state of $S$ be the maximally mixed one, $\rho_S = \mathds{1}_2/2$, thus enforcing  the time-invariance of $\omega_S\otimes\omega_E$. 
As proved in the companion paper~\cite{PRA}, the $n+1$-th step coarse-grained density matrix 
  takes the form:
\begin{align}
	\rho_S\left[{\mc{X}}^{(n+1)}\right]\ =\ \sum_{\bs{i}_{[1,n]}} \, p_{\bs{i}_{[1,n]}}\ \rho_S\left[\mc{X}_{\bs{i}_{[1,n]}}\right]\,,
	\label{cmatrix_mixture}
\end{align}
where $\rho_S\left[\mc{X}_{\bs{i}_{[1,n]}}\right]$ are sub-coarse-grained density matrices with entries 
$	\left(\rho_S\left[\mc{X}_{\bs{i}_{[1,n]}}\right]\right)_{\bs{a},\bs{b}} 
	=\omega_S\Big(\Big(X_{\bs{i}_{[1,n]}}^{\bs{b}}\Big)^\dagger X_{\bs{i}_{[1,n]}}^{\bs{a}}\Big)
	$,
where 
$
X_{\bs{i}_{[1,n]}}^{\bs{a}}:=\phi_{\bs{i}_{[1,n]}}[X_{a_{n}}]\cdots\phi_{\bs{i}_1} [X_{a_1}] X_{a_0} \in \Md{d}
$
and  $\phi_{\bs{i}_{[1,k]}}:=\phi_{i_1}\circ\,\cdots\,\circ\,\phi_{i_k}$.
Such sub-coarse-grained density matrices correspond to  closed dynamical systems. We can thus bound the entropy  of~\eqref{cmatrix_mixture} by $S\left(\rho\left[\mc{X}^{(n+1)}\right]\right)\le H\left(\pi_{[1,n]}\right)+2\log 2$, where $H(\pi_{[1,n]})=-\sum_{\bs{i}_{[1,n]}} \eta(p_{{\bs{i}}_{[1,n]}})$ and $ \eta(x)=-x\log x$. 
Then, 
the open-system ALF entropy is upper-bounded by the 
entropy rate of the chain,
\begin{equation}\label{deupperbound}
	\mathfrak{h}_S(\Theta)\
	\le\ \lim_n \,\frac{1}{n}H\left(\pi_{[1,n]}\right) =:\ \mathfrak{S}_{\omega_E}\ .
\end{equation}
Thanks to the special choice of the Pauli actions $\phi_k[X]=\sigma_kX \sigma_k$,  inequality~\eqref{deupperbound} is in fact saturated. 
Indeed, with respect to the invariant state $\mathds{1}_d / d \otimes \omega_E$ and the special POVM $\mathcal{F}$, 
the coarse-grained density matrix takes the form~\eqref{thm1_1} where the purification of $ \left({\mathds{1}_2}/{2}\right)^{\otimes n} $ yields  the maximally entangled state 
$
\ket{\psi_+^{{[1,n]}}}=2^{-n/2}\sum_{\bs{k}_{[1,n]}} \ket{\bs{k}_{[1,n]}\otimes\bs{k}_{[1,n]}}
$, while the map~\eqref{mapTn}
can be computed explicitly and amounts to an $n$-qubit Pauli channel
\begin{equation}\label{mapTnconcrete}
	\mathbb{T}_n =
	\ \sum_{\bs{i}_{[1,n]}} \,p_{\bs{i}_{[1,n]}} \,\phi_{\bs{i}_{[1,n]}}^{\otimes [1,n]} \,, \qquad \phi_{\bs{i}_{[1,n]}}^{\otimes [1,n]}:=\bigotimes_{k=1}^n \phi_{i_k}\,.
\end{equation}
The detailed derivation of~\eqref{mapTnconcrete} is carried out in the companion paper~\cite{PRA}.
Set then $\ket{\psi_{\bs{i}_{[1,n]}}}:=\left(\bigotimes_{k=1}^n \sigma_{i_k}\right)\otimes\mathds{1}_2^{\otimes n}\ket{\psi_+^{[1,n]}}$. These vectors form  a basis in $M_2^{\otimes n}(\mathbb{C})\otimes M_2^{\otimes n}(\mathbb{C})$,
so that the state
$	\mathbb{T}_n^\ddag\otimes\mathrm{id}_{d}^{\otimes n}\left[\ketbra{\psi_+^{[1,n]}}\right]=\sum_{\bs{i}_{[1,n]}}\,p_{\bs{i}_{[1,n]}} \ketbra{\psi_{\bs{i}_{[1,n]}}} $	
is diagonal with eigenvalues $\,p_{\bs{i}_{[1,n]}}$ and associated eigenvectors $\ket{\psi_{\bs{i}_{[1,n]}}}$. It follows that   inequality~\eqref{deupperbound} 
is  saturated: 
$	\mathfrak{h}_{S}(\Theta)=\mathfrak{h}_S(\Theta,\mc{F})
	=\mathfrak{S}_{\omega_E}$, 
that is, the open-system ALF  is equal to the chain mean entropy rate. In contrast to the vanishing ALF entropy of a closed finite quantum system, the previous result shows that, asymptotically, information keeps on being gathered about the open quantum system at a rate which equals that provided by the shift over the classical chain.
\par
We now let the classical environment to be a stationary Markov chain with probabilities $\pi_{[1,n]}=\big\{p_{\bs{i}_{[1,n]}}\big\}_{\bs{i}_{[1,n]}}$ forming a stationary Markov process,
$	p_{\bs{i}_{[1,n]}} = \prod_{k=2}^{n} \,T_{i_{k}\,i_{k-1}}\ p_{i_1}\,,$
defined by the stochastic matrix
$T=[T_{ij}]$, with $T_{ij}\ge0$, 
$ \sum_i T_{ij}=1$, along with $\sum_j T_{ij} \,p_j=p_i$ to ensure stationarity. The mean entropy rate of the Markov source is then equal to the two-site conditional entropy of the 
chain~\cite{Petersen_1983},
$	\mathfrak{S}_{\omega_E}=-\sum_{ij}p_j \, T_{ij} \log(T_{ij})=H(\pi_1)-I(\pi_1; \pi_2)\ ,$
where
$	
I(\pi_1;\pi_2)=H({\pi_1})+H(\pi_2)\,-\,H(\pi_{[1,2]})
$, denotes the mutual information between the first two subsequent sites of the classical chain. Because of stationarity, the  latter quantity does not depend on where the chosen pair is located along the chain and measures the correlations between any two subsequent sites. 
Concretely, choose 
\begin{gather}\label{tmatrix}
	T=\begin{pmatrix}
		p_0 & p_0 & p_0 & p_0 \\
		p   & p+\Delta& p-\Delta & p\\
		p   & p-\Delta & p+\Delta & p\\ 
		r        & r& r     & r 
	\end{pmatrix},  
\end{gather}
where  $0\le\Delta\le p\leq 1/2$, $p_0+2p+r=1$ and the invariant probability vector $\bs{p}=(p_0,p,p,r)$
that guarantees shift invariance $T\bs{p}=\bs{p}$. 
With such choices, one gets
\begin{equation}\label{specificexample}
\mathfrak{h}_{S}(\Theta)= H(\pi_{1})+2p\left[\eta(p+\Delta)+\eta(p-\Delta)-2\eta(p)\right].
\end{equation}
The associated two-site mutual information reads
$	I(\pi_{1};\pi_{2})=4 p^2 \left(\log 2- h\left(1/2+{\Delta}/{2p}\right)\right)$, 
where 
$h(x):=\eta(x)+\eta(1-x)$, $0\le x\le 1$ is the Shannon binary entropy, which is a decreasing function of $x$ for $1/2 \le x \le 1$.
\begin{figure}
	\centering
	\includegraphics[width=\linewidth]{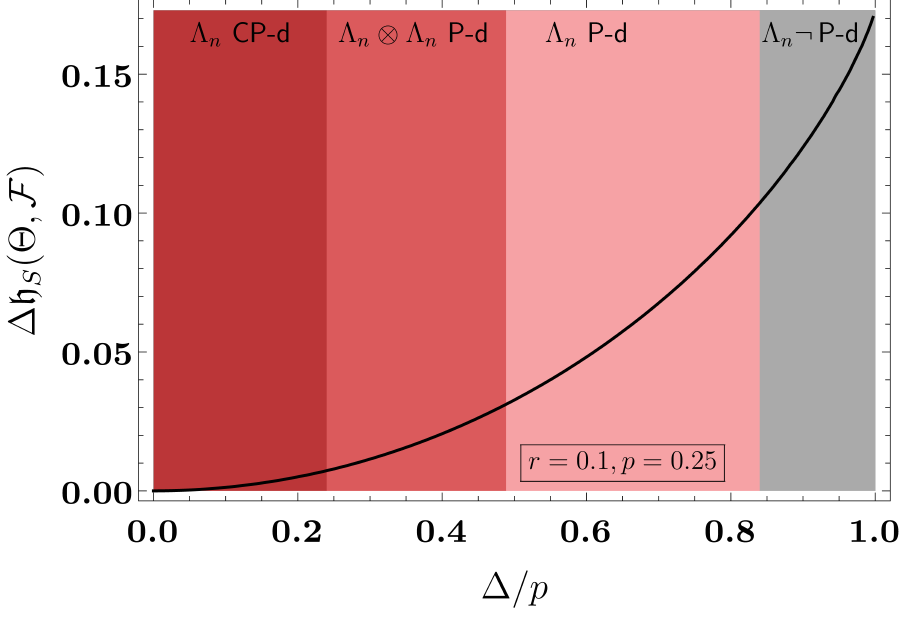}
	\caption{Entropic measure~\eqref{ineq_entropy} as a function of $\Delta/p$ for fixed values $r=0.1$ and $p=0.25$, signaling deviation from the QR regime. The colored regions correspond to different divisibility degrees of the reduced dynamics $\Lambda_n^\ddagger$. For $\Delta/p\to 1$,  $\Lambda_n^{\ddagger}$ is not P-divisible and shows distinguishability revivals; in the pink region, $\Lambda_n^\ddagger$ becomes P-divisible and does not show revivals, though  they can superactivate for  $\Lambda_n^\ddagger\otimes\Lambda_n^\ddagger$. The latter  becomes P-divisible in the red region, while in the dark-red region $\Delta/p\ll 1$, CP-divisibility is achieved.}
	\label{fig:regionsentropynew2}
\end{figure}
Therefore, the two-site correlations increase with $\Delta$ and are maximal at $p=\Delta$. Accordingly, the open-system ALF entropy is a monotonically decreasing function of the chain correlations $\Delta$.
Consider $\Delta=p=1/2$, namely the parameters for which the two site correlations are maximal. Also, from~\eqref{tmatrix} it follows that $r=p_0=0$. Correspondingly, for $p=\Delta=1/2$, one has $\mathfrak{h}_S(\Theta)=0$, 
as one would have for a closed finite system. 
It is instructive to compare such results with the behavior of the reduced dynamics of the system, which is found to be given by the  following Pauli maps
$	\Lambda_n^\ddagger=\sum_{\bs{i}_{[1,n]}} p_{\bs{i}_{[1,n]}} \, \phi_{\bs{i}_{[1,n]}}$,  for $n \ge 1$. Note that Pauli maps are self-dual: $\Lambda_n^\ddagger=\Lambda_n$.  
For the Markov chain generated by~\eqref{tmatrix}, the Schrödinger propagators $\Lambda_{m,n}^\ddag=\Lambda_n^\ddag \circ(\Lambda_{m}^{\ddag})^{-1}$ can be explicitly computed. 
One can thus evaluate the degree of memory effects of the reduced dynamics through the so-called P or CP-divisibility of the evolution,  namely, whether propagators are, respectively, positive or completely positive.
Notice that only a lack of P-divisibility detects back-flow of information in terms, for instance, of revivals of state differences due to the reduced dynamics. Indeed, lack of P-divisibility means lack of contractivity of some intertwiner $\Lambda_{n,n-1}^\ddagger$ and thus the revival of the trace distance of suitable states from time-step $n-1$ to time-step $n$~\cite{BLP,ChrusReview22}. 
In~\cite{FBGN_PhysicaScr} the divisibility features of the reduced dynamics $\Lambda_n^\ddag$ are shown to be determined by the parameter $\Delta$ as follows:
$\Lambda_n^\ddag$ is CP-divisible if and only if
$  2\Delta \le A_{p,r} \,r /p$, $
\Lambda_n^\ddag$ is P-divisible if and only if
$ 2\Delta\le  A_{p,r}\,(1+r/p)$, while the second tensor power
$ \Lambda_n^\ddag \otimes \Lambda_n^\ddag$ is P-divisible if and only if $   2\Delta \le A_{p,r} \big(1+r/p -\big(1-\sqrt{1-4p(1-2p)}\big)/2p\big)
$, where we let $A_{p,r}:=1-2(p+r)>0$.
Hence, by tuning $\Delta$ and keeping fixed the other parameters, the latter conditions establish that one can change the divisibility degree of the dynamics by increasing the correlations of the environment. 
Indeed, by increasing $\Delta$ one first loses QR, then CP-divisibility, then the P-divisibility of the second tensor power $\Lambda_n^\ddag\otimes\Lambda_n^\ddag$ and finally the P-divisibility of the one qubit dynamics $\Lambda_n^\ddag$. Note that the transitions are continuous, as already noted in literature~\cite{Megier_etal,FBGN2023,Divisibility_hierarchy}; such a behavior is compared  in Figure~\ref{fig:regionsentropynew2} with that of the entropic measure~\eqref{ineq_entropy}. 
Let us now consider the dynamical map corresponding to zero ALF entropy. 
By setting $p=\Delta=1/2$, only trivial sequences 
$\bs{i}_{[1,n]}=111\cdots $ and $\bs{i}_{[1,n]}=222\cdots $ 
are allowed by the Markov process 
with equal weights, hence yielding:
\begin{equation}\label{extreme_dyn} 
	\Lambda_n^\ddag=\begin{cases} \displaystyle\mathrm{id}_2\, & n \ \mathrm{even}\,,\\
		\displaystyle \Lambda_1\, & n \ \mathrm{odd}\,,  \qquad 
	\end{cases} \Lambda_1^\ddag[X]= \frac{1}{2}\left(\sigma_1X \sigma_1+\sigma_2X \sigma_2\right).
\end{equation}
We can explicitly study the contractivity of the trace norm under the Schrödinger dynamics $\Lambda_n^\ddag$. Since $\Lambda_1^\ddag$ is CPTP, the trace norm is always contractive between step $2m$ and $2m+1$.
In particular, take $X=X^\dagger=\sum_{k=0}^3 x_k \,\sigma_k$, $x_k\ne 0$, with $x_0< \norm{\bs{x}}=\sqrt{x_1^2+x_2^2+x_3^2}$, so that $\normt{X}=2 \norm{\bs{x}}$ and $\|{\Lambda_1^\ddag[X]}\|_1=2 |x_3| $. 
Then, 
$	\|{\Lambda^\ddag_{2m+1}[X]}\|_1-\norm*{\Lambda^\ddag_{2m}[X]}_1=\|{\Lambda^\ddag_{1}[X]}\|_1-\normt{X}\nonumber=-2(\norm{\bs{x}}-\abs{x_3}) <0$, $m \ge0$.
Between steps $2m-1$ and $2m$, $m\ge1$, instead, we have the converse inequality:
$ 
\|{\Lambda^\ddag_{2m}[X]}\|_1-\|{\Lambda^\ddag_{2m-1}[X]}\|_1=\normt{X}-\|{\Lambda^\ddag_{1}[X]}\|_1> 0
$, 
so that the trace norm always revives at odd times, signaling back-flow of information. 
According to the operational interpretation discussed above,
	a low, possibly zero, ALF entropy rate measures a  flow of information back into the open quantum system $S$. Indeed,  asymptotically, less and less uncertainty about the outcome of the next measurement performed on $S$ is left by the collisional dynamics. In particular, 
	 a vanishing dynamical entropy indicates that, asymptotically, no new information can be extracted by further probing the open system. 
 Remarkably, the dynamical map~\eqref{extreme_dyn} exhibits an extreme back-flow of information since the trace norm revives after each odd time. Accordingly, the open-system ALF entropy is zero, as for a closed system dynamics. 
\par
For the model at hand, QR-Markovianity holds if and only if the dissipative dynamics is a semigroup, namely for $\Delta=0$, that is in the case of an uncorrelated classical environment.
 Nonetheless, as emerges from the previous discussion,  $\Lambda_n^\ddagger$ can be P-divisible or even CP-divisible with $\Delta>0$. In such cases, the reduced dynamics alone cannot detect memory effects through trace-distance revivals. 
Instead, memory effects can be assessed at a deeper level.   
In this regard,~\eqref{GNS_matrix} suggests to define the following CPTP map acting on the GNS doubling of the open system
\begin{equation}\label{GNS_diss}
	\Gamma_n^{\mathbb{X}}[\bullet]:=\Tr_{E}\!\left( \Big(\mathbb{U}_{\Theta}^\ddag\circ \big(\mathbb{X}^\ddag\otimes\mathrm{id}_d\otimes\mathrm{id}_E\big)\Big)^n\!\left[\bullet\otimes \ketbra{\Omega_E} \right]\right),
\end{equation}
 which parametrically depends on the measurements. 
Switching off the measurements ($\mathbb{X}=\mathrm{id}$), $\Gamma_n^{\mathrm{id}}$ describes the dissipative dynamics in the GNS representation. 
For the qubit model with dynamics generated by the classical Markov chain environment~\eqref{tmatrix}, we then have that  
$
\Lambda_n^\ddagger$ is CP-divisible  if and only if  $\Gamma_n^{\mathrm{id}}$ is P-divisible.
As a consequence, if $\Lambda_n^\ddagger$ is P-divisible, it is contractive and cannot exhibit back-flow of information as revivals of the trace distances. On the other hand, if in addition $\Lambda_n^\ddagger$ is not CP-divisible, then
the dilated map $\Gamma_n^{\mathrm{id}}$ to two qubits cannot be P-divisible and thus detects back-flow of information at the higher level of a GNS doubling of the qubit dynamics.\\
\indent To summarize, we studied the  open-system ALF entropy for a concrete collisional model whereby it provides the maximal amount of information that can be extracted from the dissipative dynamics by repeated POVM measurements on the open system. 
The lower the ALF entropy, the lower the information retrievable, asymptotically, by repeated measurements.
For these reasons, we propose the ALF entropic measure~\eqref{ineq_entropy} as a stronger measure of back-flow of information accounting for the multi-time statistics of the process. The latter can detect memory effects that cannot be assessed by the reduced dynamics alone. This is particularly evident in our qubit example, whereby the proposed measure is positive while the reduced dynamics may be P- or even CP-divisible (see Fig.~\ref{fig:regionsentropynew2}). The same behavior emerges from the GNS representation. In this context, the bipartite dissipative dynamics~\eqref{GNS_diss} represents the marginal of the GNS system–ancilla–environment evolution, whose information content is encoded in the ALF entropy. Again,  non-Markovian effects emerge that are hidden when inspecting  the reduced dynamics only.

For a qubit coupled to  
a Markov chain environment, the open-system ALF entropy decreases with increasing next-neighbor correlations since memory effects also increase.  
A singular instance where the entropic and the reduced dynamics points of view coincide is provided by a dissipative Pauli qubit dynamics with periodic trace-norm revivals and vanishing ALF entropy, as much as for a closed qubit
time-evolution. \par
The authors acknowledge financial support from PNRR MUR project PE0000023-NQSTI.

\bibliographystyle{apsrev4-2}

%


\end{document}